\documentclass[showpacs,pre,superscriptaddress,twocolumn]{revtex4}
\usepackage{amssymb}
\usepackage{graphicx}
\usepackage{amsmath}
\usepackage{amsfonts}

\begin{document}

\title{The Application of Asymmetric Entangled States in Quantum Game}
\author{Ye Li}
\affiliation{Hefei National Laboratory for Physical Sciences at
Microscale and Department of Modern Physics, University of Science
and Technology of China, Hefei 230026, People's Republic of China}
\author{Gan Qin}
\affiliation{Hefei National Laboratory for Physical Sciences at
Microscale and Department of Modern Physics, University of Science
and Technology of China, Hefei 230026, People's Republic of China}
\author{Jiangfeng Du}
\email{djf@ustc.edu.cn} \affiliation{Hefei National Laboratory for
Physical Sciences at Microscale and Department of Modern Physics,
University of Science and Technology of China, Hefei 230026,
People's Republic of China} \affiliation{Department of Physics,
National University of Singapore, Lower Kent Ridge, Singapore
119260, Singapore}

\begin{abstract}
In the present letter, we propose a more general entangling
operator to the quantization of Cournot economic model, in which
players can access to a continuous set of strategies. By analyzing
the relation between the von Neumann entropy of the entangled
state and the total profit of two players precisely, we find that
the total profit at the Nash equilibrium always achieves its
maximal value as long as the entropy tends to infinity. Moreover,
since the asymmetry is introduced in the entangled state, the
quantum model shows some kind of "encouraging" and "suppressing"
effect in profit functions of different players.
\end{abstract}

 \maketitle

\section{Introduction}

Game theory has experienced a great development since early time
of last century through its wide application in nearly every
aspect of our modern society. In recent few years, along with the
thriving of quantum information, a newly-emerging field, quantum
game has begun to attract much attention due to the intimate
connection between the theory of game and the theory of quantum
communication \cite{Eisert}. This field was initiated by a paper
of Meyer, finding that a player can always beats his classical
opponent by adopting quantum strategies on the coin tossing game
\cite{Meyer}. Later, Eisert \textit{et al.} \cite{Eisert}
introduced quantum entanglement into the famous Prisoners'
Dilemma, by which the contradiction existing in the classical
scenario vanishes. Since then, quantum game theory has developed
rapidly and exhibited great superiority over its classical
counterpart in later works \cite{3,4,5,6,7,8,9,10,11,12,13}. In
addition to theoretical investigations, Du \textit{et al.}
realized the quantum Prisoners' Dilemma in experiments for the
first time \cite{14}, and therefore constructed a practical base
for quantum game.

However, most of previous investigations on quantum game mainly
focused on game models with discrete strategies set. In Li
\textit{et al.}'s paper \cite {Li}, they first put forward a
quantization scheme for Cournot duopoly, a famous economical model
even preceding the birth of game theory \cite{Cournot}, in which
two players can virtually cooperate and get the optimal profit at
the maximal quantum entanglement between them albeit both act
"selfishly" just as they do in the classical case. This work
actually set up a foundation for further discussions regarding
quantum game with continuous strategies \cite{17,18,19}. Recently
Qin \textit{et al.} introduced a quantization scheme for
asymmetric games \cite{20}. The key modification of their scheme
compared with Li \textit{et al.}'s is the adoption of two
different entanglement factors, which ensure the asymmetric games
to get the optimal cooperative profit. Nevertheless, Qin
\textit{et al.}'s quantum scheme can still be regarded as a fair
play, for the two factors are determined by the extent of the
asymmetry of the game. It is quite often that various environments
may affect the quantum game, leading to asymmetric forms of the
entanglement operator.

Hence, to better understand the property of quantum game, it's
worthwhile studying the distinctive features of quantum game
resulted from the change of the entanglement operator. In this
letter we develop the quantum model of Cournot's duopoly proposed
by Li \textit{et al.} by changing the initial state before
actually entangling them, and we define these operations
altogether as a whole entangling operation. Thus it can be
discovered that the total profit at the unique Nash equilibrium
will increase monotonously as long as the entropy of the entangled
state, namely the entanglement, tends to infinity. While,
considering the lack of symmetry in this quantization scheme, our
result will not remain Pareto optimal (the best result that can be
achieved without disadvantaging at least one group), which in fact
may not be a necessity in the real world because of the simple
fact that there is no absolutely-impartial law. Besides, all of
the operators we cite in the quantization of Cournot duopoly can
be realized experimentally within the capacity of modern optical
technology through proper design.

\section{Cournot Duopoly}

Now we briefly recall the classical Cournot duopoly. Duopoly is an economic
condition in which two firms hold the market of a certain commodity without
a third competitor. In Cournot model, two firms are assumed to produce a
homogeneous product and simultaneously decide their quantities $q_{1}$ and $%
q_{2}$. and the price of this product is decided by the quantities as

\begin{equation}
p=\left\{
\begin{array}{ccc}
a-Q & \text{ } & (Q<a), \\
0 & \text{ } & (Q\geq a),%
\end{array}
\right.
\end{equation}
where $Q$ is the total quantity of two firms, i.e. $Q=q_{1}+q_{2}$ and $a$
is a constant. Assume that the unit cost of the product is also a constant $%
c $, the profit function of each firm would be:

\begin{equation}
U_{j}=q_{j}(p-c)=q_{j}(k-(q_{1}+q_{2})),  \label{cl-profit}
\end{equation}
with $k=a-c>0$ and $j=1,2$. It's easy to work out the unique Nash
equilibrium:

\begin{align}
q_{1}^{\ast }& =q_{2}^{\ast }=\frac{k}{3},  \notag \\
u_{1}^{\ast }& =u_{2}^{\ast }=\frac{k^{2}}{9}.
\end{align}

However, the Nash equilibrium is not the optimal solution for the
market. If the two firms cooperate and raise their total quantity
to $k/2$, they can acquire \emph{the highest total profit
}$k^{2}/4$. In addition, by keeping the symmetry of the model, we
can obtain the Pareto optimal in the following solution:

\begin{align}
q_{1}^{\ast }& =q_{2}^{\ast }=\frac{k}{4},  \notag \\
u_{1}^{\ast }& =u_{2}^{\ast }=\frac{k^{2}}{8}.  \label{cl-co-result}
\end{align}

The profit difference between the optimal choice and the Nash
equilibrium reflects the conflict between individual rationality
and collective rationality, just like in the classical prisoners'
dilemma.

\section{Quantization of Classical Model}

\begin{figure}[ptb]
\begin{center}
\includegraphics[
height=3 cm, width=0.5\textwidth ]{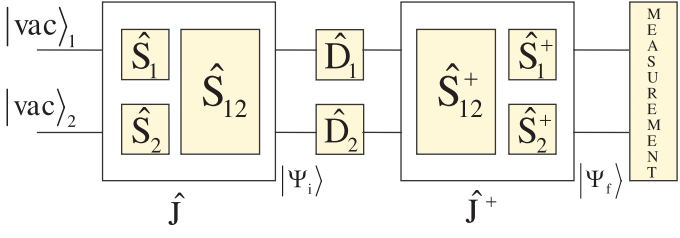}
\end{center}
\caption{The quantum structure of the game.}
\label{structure}
\end{figure}

Here we put forward a modified quantization model of Li \textit{et
al.}'s \cite{Li}, where we make use of two single-mode
electromagnetic fields. Fig. \ref{structure} shows the quantum
structure of the game. As a extension, we first apply two
single-mode squeezing operator $\hat{S}_{1}\otimes \hat{S}$ to the
vacuum state, where $\hat{S}_{j}=\exp (\frac{\gamma _{j}}{2}a_{j}{}^{2}-%
\frac{\gamma _{j}}{2}a_{j}^{\dag }{}^{2})$, and $a_{j}^{\dag
},a_{j}$ are the creation and annihilation operator of the
corresponding field, followed by a two-mode squeezing operator
$\hat{S}_{12}=\exp (\gamma a_{1}a_{2}-\gamma a_{1}^{\dag
}a_{2}^{\dag })$ of a two-mode electromagnetic field. Thus the
total entanglement operator would be:

\begin{align}
\hat{J}(\gamma _{1},\gamma _{2},\gamma _{12}) & = \hat{S}_{12}\hat{S}_{1}\hat{%
S}_{2} \notag \\
& = \exp [\gamma _{12}(a_{1}a_{2}-a_{1}^{\dag }a_{2}^{\dag })] \notag \\
& \cdot \exp [\gamma _{1}(a_{1}{}^{2}-a_{1}^{\dag }{}^{2})/2] \notag \\
& \cdot \exp [\gamma _{2}(a_{2}{}^{2}-a_{2}^{\dag }{}^{2})/2],
\end{align}
with constraint $\gamma _{1},\gamma _{2},\gamma _{12}\in R$ and
$\gamma _{12}\geq 0$ (which also is an extension of Li \textit{et
al}'s model).
Apparently, we can turn back to Li \textit{et al}'s entanglement operator if $%
\gamma _{1}=\gamma _{2}=0$. The strategies in the set of local unitary
operators is $\hat{D}_{j}(x_{j})=\exp [x_{j}(a_{j}^{\dag }-a_{j})/\sqrt{2}]$
as the scheme of Li \textit{et al.}'s.

The final measurement is made corresponding to the observables $\hat{X}%
_{j}=(a_{j}^{\dag }+a_{j})/\sqrt{2}$, which in quantum optics is the
amplitude quadrature (we can simply perceive it as the "position" operator).
This measurement is done by the homodyne measurement assuming that the state
is infinitely squeezed. Correspondingly, by citing the phase quadrature $%
\hat{P}_{j}=i(a_{j}^{\dag }-a_{j})/\sqrt{2}$ (which can be seen as
the "momentum" operator), we can rewrite the entanglement operator
and the strategies in the following form:

\begin{align}
\hat{J}(\gamma _{1},\gamma _{2},\gamma _{12})& =\exp [i\gamma _{12}(\hat{X}%
_{1}\hat{P}_{2}+\hat{X}_{2}\hat{P}_{1})]  \notag \\
& \quad \,\cdot \exp [i\gamma _{1}(\hat{X}_{1}\hat{P}_{1}+\hat{P}_{1}\hat{X}%
_{1})/2]  \notag \\
& \quad \,\cdot \exp [i\gamma _{2}(\hat{X}_{2}\hat{P}_{2}+\hat{P}_{2}\hat{X}%
_{2})/2], \\
\hat{D}_{j}(x_{j})& =\exp (-ix_{j}\hat{P}_{j}),\qquad j=1,2.
\end{align}

Obviously, the classical Cournot duopoly is only a special case with $\gamma
_{12}=0$ in its quantum extension. Because the strategy set is not extended
compared to its classical counterpart, we can be sure that all the features
shown in the quantization version attribute to the entanglement of the
states. \emph{The specific property of the entangled state is analyzed in
the paragraphs below.}

By detailed calculation we get:

\begin{align}
\hat{J}^{\dag }\hat{D}_{1}(x_{1})\hat{J}& =\exp [-ix_{1}(\hat{P}
_{1}e^{\gamma _{1}}\cosh \gamma _{12}+\hat{P}_{2}e^{\gamma _{2}}\sinh \gamma
_{12})], \\
\hat{J}^{\dag }\hat{D}_{2}(x_{2})\hat{J}& =\exp [-ix_{2}(\hat{P}
_{2}e^{\gamma _{2}}\cosh \gamma _{12}+\hat{P}_{1}e^{\gamma _{1}}\sinh \gamma
_{12})].
\end{align}

Therefore the final state reads:

\begin{align}
\left\vert \psi _{f}\right\rangle & =\hat{J}^{\dag }(\hat{D}_{1}\otimes \hat{%
D}_{2})\left\vert \psi _{i}\right\rangle  \notag \\
& =\exp [-ie^{\gamma _{1}}(x_{1}\cosh \gamma _{12}+x_{2}\sinh \gamma _{12})%
\hat{P}_{1}]\left\vert vac\right\rangle _{1}  \notag \\
& \otimes \exp [-ie^{\gamma _{2}}(x_{2}\cosh \gamma _{12}+x_{1}\sinh \gamma
_{12})\hat{P}_{2}]\left\vert vac\right\rangle _{2}.
\end{align}

And the measurement gives the respective quantity of each firm
player:

\begin{align}
q_{1}& =e^{\gamma _{1}}(x_{1}\cosh \gamma _{12}+x_{2}\sinh \gamma _{12}), \\
q_{2}& =e^{\gamma _{2}}(x_{2}\cosh \gamma _{12}+x_{1}\sinh \gamma _{12}).
\end{align}

Simply by substituting the above into the classical profit functions (see Eq. (\ref%
{cl-profit})), we get the quantum profit:

\begin{align}
u_{1}^{Q}(x_{1},x_{2})& =e^{\gamma _{1}}(x_{1}\cosh \gamma _{12}+x_{2}\sinh
\gamma _{12})p(x_{1},x_{2}), \\
u_{2}^{Q}(x_{1},x_{2})& =e^{\gamma _{2}}(x_{2}\cosh \gamma _{12}+x_{1}\sinh
\gamma _{12})p(x_{1},x_{2}),
\end{align}
where $p(x_{1},x_{2})=[k-x_{1}(e^{\gamma _{1}}\cosh \gamma
_{12}+e^{\gamma _{2}}\sinh \gamma _{12})-x_{2}(e^{\gamma
_{2}}\cosh \gamma _{12}+e^{\gamma _{1}}\sinh \gamma _{12})]$.

Solving the Nash equilibrium equations we obtain the unique
solution:

\begin{align}
x_{1}^{\ast }& =\frac{ke^{\gamma _{2}}\cosh \gamma _{12}}{(e^{\gamma
_{1}}+e^{\gamma _{2}})\sinh 2\gamma _{12}+e^{\gamma _{1}+\gamma _{2}}(2\cosh
2\gamma _{12}+1)}, \\
x_{2}^{\ast }& =\frac{ke^{\gamma _{1}}\cosh \gamma _{12}}{(e^{\gamma
_{1}}+e^{\gamma _{2}})\sinh 2\gamma _{12}+e^{\gamma _{1}+\gamma _{2}}(2\cosh
2\gamma _{12}+1)}.
\end{align}

And the profit at the equilibrium reads:

\begin{widetext}
\begin{align}
u_{1}^{Q}(x_{1}^{\ast },x_{2}^{\ast })& =\frac{\cosh \gamma _{12}(\cosh
\gamma _{12}+e^{\Delta \gamma }\sinh \gamma _{12})^{2}(\cosh \gamma
_{12}+e^{-\Delta \gamma }\sinh \gamma _{12})}{(1+2\cosh 2\gamma _{12}+2\cosh
\Delta \gamma \sinh 2\gamma _{12})^{2}}k^{2}, \\
u_{2}^{Q}(x_{1}^{\ast },x_{2}^{\ast })& =\frac{\cosh \gamma _{12}(\cosh
\gamma _{12}+e^{-\Delta \gamma }\sinh \gamma _{12})^{2}(\cosh \gamma
_{12}+e^{\Delta \gamma }\sinh \gamma _{12})}{(1+2\cosh 2\gamma _{12}+2\cosh
\Delta \gamma \sinh 2\gamma _{12})^{2}}k^{2},
\end{align}
\end{widetext}
where $\Delta \gamma =\gamma _{1}-\gamma _{2}$. Apparently, the
profit functions only depend on two parameters $\gamma _{12}$ and
$\Delta \gamma $.
In addition, if we make $\gamma _{1}=\gamma _{2}=0$, i.e. $\Delta \gamma =0$%
, it's easy to find that the result is exactly that of Li \textit{et al} in
their quantization scheme.

\section{Entropy \& Asymmetry}

Using the method put forward in Rendell's paper \cite{Rendell}, we write the
entangled state in the representation of "position":

\begin{equation}
\left\vert \psi _{i}\right\rangle =\exp [-(\alpha x_{1}^{2}+\beta
x_{2}^{2}+2\gamma x_{1}x_{2})/2].
\end{equation}

By solving the equations below:

\begin{align}
\hat{J}a_{1}\hat{J}^{\dag }\left\vert \psi _{i}\right\rangle & =\hat{J}a_{1}%
\hat{J}^{\dag }\hat{J}\left\vert vac\right\rangle _{1}=0, \\
\hat{J}a_{2}\hat{J}^{\dag }\left\vert \psi _{i}\right\rangle & =\hat{J}a_{2}%
\hat{J}^{\dag }\hat{J}\left\vert vac\right\rangle _{2}=0.
\end{align}

We can obtain the expression of the three unknown parameters as

\begin{align}
\alpha & =\frac{(1+\lambda _{1})(1-\lambda _{2})+(1-\lambda _{1})(1+\lambda
_{2})\lambda _{12}^{2}}{(1-\lambda _{1})(1-\lambda _{2})(1-\lambda _{12}^{2})%
}, \\
\beta & =\frac{(1-\lambda _{1})(1+\lambda _{2})+(1+\lambda _{1})(1-\lambda
_{2})\lambda _{12}^{2}}{(1-\lambda _{1})(1-\lambda _{2})(1-\lambda _{12}^{2})%
}, \\
\gamma & =\frac{2\lambda _{12}(1-\lambda _{1}\lambda _{2})}{(1-\lambda
_{1})(1-\lambda _{2})(1-\lambda _{12}^{2})},
\end{align}
where $\lambda _{1}=\tanh \gamma _{1},\lambda _{2}=\tanh \gamma
_{2},\lambda _{12}=\tanh \gamma _{12}$. To better assess the
entanglement of the state, we need to calculate the von Neumann
entropy of the state, which is defined
as $S=-$Tr$_{1}(\rho _{1}\ln \rho _{1})=-$Tr$_{2}(\rho _{2}\ln \rho _{2})$ ($%
\rho _{1},\rho _{2}$ are the density matrix of two single-mode
electromagnetic fields). Based on the above calculation we get the following
equality by making use of Rendell's method:

\begin{equation}
S(\eta )=\ln (\eta /2)+\frac{1}{2}\sqrt{\eta ^{2}+1}\ln
\frac{\sqrt{\eta ^{2}+1}+1}{\sqrt{\eta ^{2}+1}-1}, \label{entropy}
\end{equation}
where $\eta =\sinh 2\gamma _{12}\cosh \Delta \gamma \geq 0$. As a
confirmation, when $\gamma _{1}=\gamma _{2}=0$, Eq.
(\ref{entropy}) is reduced to:

\begin{equation}
S=\cosh ^{2}\gamma _{12}\ln (\cosh ^{2}\gamma _{12})-\sinh ^{2}\gamma
_{12}\ln (\sinh ^{2}\gamma _{12}),
\end{equation}
which is exactly the entropy of a two-mode squeezing state. It is
interesting to find that like the profit function, the entropy is
also determined by only two parameters $\gamma _{12}$ and $\Delta
\gamma $. In addition, we find that the von Neumann entropy
monotonously increases from zero to infinity as the parameter
$\eta $ increases. Since $\eta $ increases as either of the two
parameters increases (when $\gamma _{12}\neq 0$), the
entropy is an increasing function of both $\gamma _{12}$ and $\Delta \gamma $%
.

In Li \textit{et al.}'s entanglement operator, through \emph{whole
unitary transformation} $\exp [\gamma _{12}(a_{1}a_{2}-a_{1}^{\dag }a_{2}^{\dag })]$%
, we get the maximal entangled state as $\gamma _{12}$ tends to
infinity. Nevertheless, it's discovered that only by changing the
initial state via \emph{local unitary transformation} $\exp
[\gamma _{i}(a_{i}{}^{2}-a_{i}^{\dag }{}^{2})/2]$, the maximized
entanglement is also achieved as long as entanglement exists (i.e.
$\gamma _{12}\neq 0$, even if it is very small; see Fig.
\ref{r12-fixed}). Obviously, such kind of situation is impossible
in quantum game with Hilbert space of finite dimensions.

Moreover, we estimate the asymmetry by calculating the relative difference
of the "position" uncertainties (which actually represent the width of
Gaussian wave package in each "position" space) between two parties in the
expression below:

\begin{equation}
\frac{\left\langle \Delta x_{1}^{2}\right\rangle -\left\langle \Delta
x_{2}^{2}\right\rangle }{\left\langle \Delta x_{1}^{2}\right\rangle
+\left\langle \Delta x_{2}^{2}\right\rangle }=-\frac{\tanh \Delta \gamma }{%
\cosh 2\gamma _{12}}.
\end{equation}

For fixed and finite $\gamma _{12}$, the above expression would monotonously
decreases as $\Delta \gamma $ increases. Due to the obvious fact that the
system is completely symmetric when $\Delta \gamma =0$, we can simply
perceive $\left\vert \Delta \gamma \right\vert $ as a measurement of the
asymmetry of the entangled state, while it's worth mentioning that all the
discussion concerning asymmetry above only reflects a variation trend
between variables and the final state.

\section{Discussion}

So far, we have confined three parameters presented in the
entanglement operator to only two ones, i.e. $\gamma _{12},\Delta
\gamma $. In the paragraphs below, specific analysis of the
relation between the game model adopted here and the entangled
state is given.

\begin{figure}[tbp]
\begin{center}
\includegraphics[
height=3.85in, width=3.1in ]{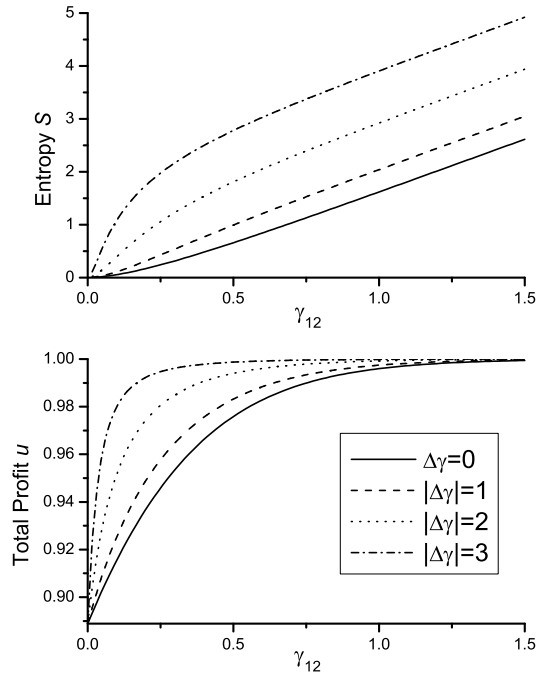}
\end{center}
\caption{when $\Delta \protect\gamma $ is fixed, the total profit
$u$ compares to the entropy $S$ as $\protect\gamma _{12}$
changes.} \label{dr-fixed}
\end{figure}

\begin{figure}[tbp]
\begin{center}
\includegraphics[
height=3.9in, width=3in]{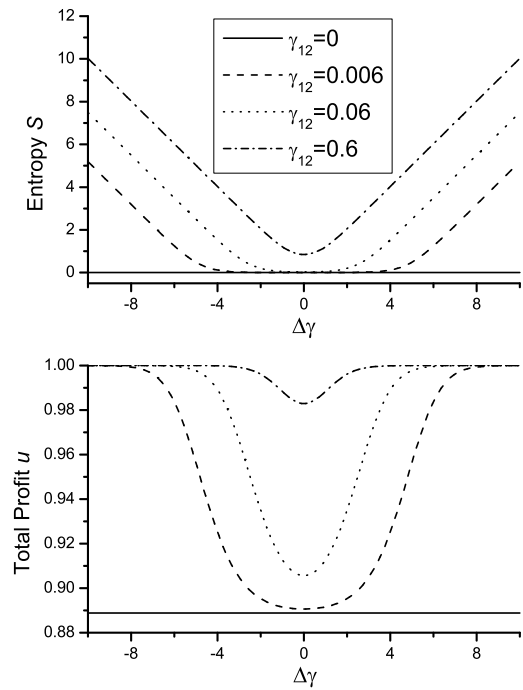}
\end{center}
\caption{when $\protect\gamma _{12}$ is fixed, the total profit
$u$ compares to the entropy $S$ as $\Delta \protect\gamma $
changes.} \label{r12-fixed}
\end{figure}

Since the total profit of two players cannot achieve the maximum
at the Nash equilibrium due to the deviation between individual
rational choice and collective optimization requirement in
classical Cournot duopoly, it must be the application of quantum
entanglement that eliminates this dilemma in Li \textit{et al.}'s
scheme. As an extension of their entangled state, we separately
give the alternation trends of the total profit ($u=u_{1}+u_{2}$)
and the entropy $S$ versus $\gamma _{12},\Delta \gamma $ in the
following figures (Fig. \ref{dr-fixed}, Fig. \ref{r12-fixed}), in
which the unit of profit is taken as $k^{2}/4$ , namely, the
optimal value of the total profit in the classical cooperation
scenario (See Eq. (\ref{cl-co-result})).

It's obvious to see that both the profit and entropy function increases as
either of the parameters increases (the condition $\gamma _{12}\neq 0$ is
necessary here; otherwise there would be no entanglement between two
parties). When the entropy tends to infinity, the total profit also achieves
its maximal value, and Li \textit{et al.}'s situation is just a special case
when $\Delta \gamma =0$. Since the entropy is defined as a measurement of
the entanglement of the state, it would be adequate to explain that the
entanglement of the state does help improve the total profit in Cournot
duopoly independent of the concrete form of the state (i.e. no matter in
which way the entanglement is increasing) as illustrated in the previous
section, and so long as \emph{the maximal entanglement} is attained, the
conflict between individual rationality and collective rationality
disappears.

\begin{figure}[tbp]
\begin{center}
\includegraphics[
height=2.4in, width=0.47\textwidth]{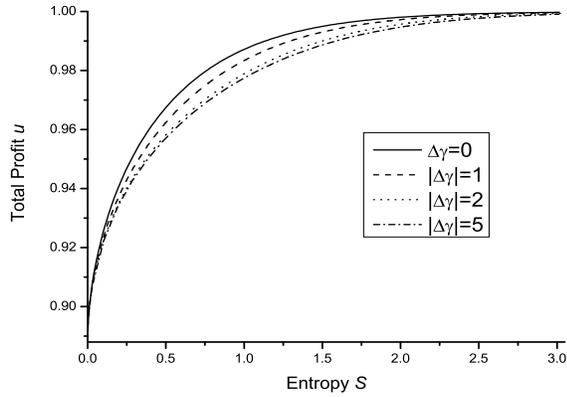}
\end{center}
\caption{when $\Delta \protect\gamma $ is fixed, the function plot between
the entropy $S$ and total profit $u$.}
\label{s-u}
\end{figure}

In Fig. \ref{s-u}, we present a more detailed function plot
between the entropy and the total profit. Apparently, although
both quantities have the same increasing trend as seen in Fig.
\ref{dr-fixed} and Fig. \ref{r12-fixed}, the entropy does not
determine the total profit uniquely, i.e. there isn't a one-to-one
correspondence between such two quantities. In combination of
previous discussion concerning asymmetry, it is not difficult to
find that \emph{the increase of asymmetry} would cause the
reducing of the total profit if the entanglement of the state
don't reach the maximum.

\begin{figure}[ptb]
\begin{center}
\includegraphics[
height=2.5in, width=0.5\textwidth]{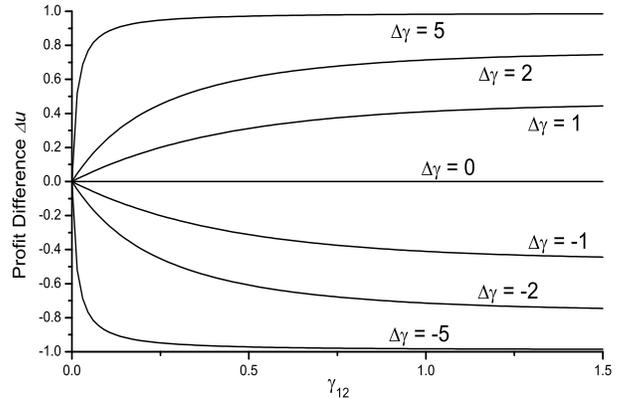}
\end{center}
\caption{the variation trend of the profit difference as
parameters change.} \label{effect-of-dr}
\end{figure}

Furthermore, the variation of asymmetry also results in the change
of the profit difference between two players in our quantization
scheme. Fig. \ref{effect-of-dr} shows
that Player 1 would monopolize the market and get the optimal profit when $%
\Delta \gamma $ tends to positive infinity, while on the contrary,
player 2 would become the monopolizer of the market if $\Delta
\gamma $ tends to negative infinity. Therefore, $\gamma
_{1},\gamma _{2}$ can be separately regarded as the extent to
which player 1 and 2 are encouraged (or the corresponding
opponents are suppressed), that is, the larger the parameter is,
the more profit the corresponding player can acquire.

\section{Conclusion}

We investigate the quantization of games with continuum strategic
space by making use of a more general entanglement operator. For
the particular case of symmetric scenario, it would turn back to
the quantization scheme put forward by Li \textit{et al.}. As an
extension of their quantum model, we investigate \emph{the
relation between the Nash equilibrium in quantum situation and the
entanglement of the state} more specifically, and find that the
optimal total profit can always be attained as long as the maximal
entanglement is realized. Also we observed some novel features
such as the encouraging and suppressing effect in our quantum
model, which completely attributes to \emph{the asymmetry of the
system}. Therefore, in our quantization scheme, the government
would have more choices in managing the market. It actually
constitutes a better base for further discussion of game models
involving incomplete information or unequal competition status.

\section{Acknowledgements}

This work is supported by the National fundamental Research
Program (Grant No. 2001CB309300), the National Science Fund for
Distinguished Young Scholars (Grant No. 10425524) and the Project
in Quantum Information Technology (Grant No. R-144-000-071-305).


\begin{thebibliography}{99}
\bibitem{Eisert} J. Eisert, M. Wilkens, M. Lewenstein, Phys. Rev. Lett. 83
(1999) 3077.

\bibitem{Meyer} D.A. Meyer, Phys. Rev. Lett. 82 (1999) 1052.

\bibitem{3} L. Marinatto, T. Weber, Phys. Lett. A 272 (2000) 291.

\bibitem{4} N.F. Johnson, Phys. Rev. A 63 (2001) 020302.

\bibitem{5} W.Y. Hwang, D. Ahn, S.W. Hwang, Phys. Rev. A 64 (2001) 064302.

\bibitem{6} J.L. Chen, L.C. Kwek, C.H. Oh, Phys. Rev. A 65 (2002) 052320.

\bibitem{7} A. Iqbal, A.H. Toor, Phys. Rev. A 65 (2002) 022306; Phys. Rev. A
65 (2002) 052328; J. Phys. A: Math. Gen. 37 (2004) 5873.

\bibitem{8} A.P. Flitney, D. Abbott, Phys. Rev. A 65 (2002) 062318; J. Phys.
A: Math. Gen. 38 (2005) 449.

\bibitem{9} H. Fort and S. Viola, Phys. Rev. E 69 (2004), 036110.

\bibitem{10} A. Nawaz and A. H. To, J. Phys. A: Math. Gen. 37 (2004) 4437.

\bibitem{11} S.K. Oumlzdemir, J. Shimamura, N. Imoto, Phys. Lett. A 325
(2004) 104; Shimamura J, Ozdemir SK, Morikoshi F, Imoto N, Int. J. Quan.
Inform. 2 (2004) 79.

\bibitem{12} E.W. Piotrowski, J. Sladkowski, 2 (2004) 495; Physica A, 345
(2005) 185.

\bibitem{13} Q. Chen, Y. Wang, J.T. Liu, K.L. Wang, Phys. Lett. A 327 (2004)
98.

\bibitem{14} J. Du, H. Li, X. Xu, M. Shi, J. Wu, X. Zhou, R. Han, Phys. Rev.
Lett. 88 (2002) 137902.

\bibitem{Cournot} A. Cournot, Researches Into the Mathematical Principles of
the Theory of wealth, Macmillan, New York, 1897.

\bibitem{Li} H. Li, J. Du, S. Massar, Phys. Lett. A 306 (2002) 73-78.

\bibitem{17} J.F. Du, H. Li, C.Y. Ju, Phys. Rev. E 68 (2003) 016124; J.F.
Du, H. Li, C.Y. Ju, J. Phys. A: Math. Gen. 38 (2005) 1559.

\bibitem{18} C.F. Lo, D. Kiang, Phys. Lett. A 318 (2003) 333, Phys. Lett. A
321 (2004) 94, Phys. Lett. A 346 (2005) 65.

\bibitem{19} G. Qin, X. Chen, M. Sun, J.F. Du, J. Phys. A: Math. Gen. 38
(2005) 4247; X. Chen, G. Qin, X.Y. Zhou, J.F. Du, Chin. Phys. Lett. 22
(2005) 1033.

\bibitem{20} G. Qin, X. Chen, M. Sun, X.Y. Zhou, J.F. Du, Phys. Lett. A 340
(2005) 78.

\bibitem{Rendell} R.W. Rendell and A.K. Rajagopal, Phys. Rev. A 72, 012330 (2005).

\bibitem{22} D.F. Walls and G.J. Milburn, Quantum Optics, Springer-Verlag.

\bibitem{23} S.L. Braunstein, Review of Modern Physics, 77, 513 (2005).

\bibitem{24} S.L. Braunstein and H.J. Kimble, Phys. Rev. Lett. 80, 869 (1998).
\end{thebibliography}
\end{document}